\documentclass[aps,pra,twocolumn]{revtex4-2}

\usepackage{graphicx}
\usepackage{amsmath,amssymb,amsfonts}
\usepackage{bm}

\begin{document}

\title{Map-Dependent Quantum Characteristic Functions and CP-Divisibility in Non-Markovian Quantum Dynamics}

\author{Koichi Nakagawa}
\affiliation{Hoshi University}

\begin{abstract}

We introduce map-dependent quantum characteristic functions constructed from the normalized Choi operator of quantum dynamical maps. 
We prove a Bochner--Choi positivity theorem establishing that the positive-type condition of the associated Gram matrix is equivalent to complete positivity of the underlying quantum channel. 
Applying the construction to intermediate dynamical maps, we obtain a characterization of CP-divisibility in terms of positivity of two-time characteristic functions. 
Numerical examples for amplitude damping and pure dephasing models demonstrate that negativity of the Gram matrix coincides with the breakdown of CP-divisibility and the emergence of information backflow. 
The proposed framework provides a new bridge between characteristic-function methods in quantum statistics and structural properties of quantum dynamical maps.

\end{abstract}

\maketitle

\section{Introduction}

Understanding memory effects in open quantum systems has become a central problem in quantum physics.
Non-Markovian dynamics arise when correlations between a system and its environment persist over time, leading to deviations from semigroup evolution.

Several approaches to quantifying non-Markovianity
have been proposed. One prominent approach is the
information backflow measure introduced by Breuer,
Laine, and Piilo (BLP) \cite{Breuer2009}, based on the
revival of distinguishability between quantum states.
Another approach is the divisibility criterion proposed
by Rivas, Huelga, and Plenio (RHP) \cite{Rivas2010},
which characterizes Markovian dynamics through the
complete positivity of intermediate dynamical maps.

The structure of dynamical maps and their divisibility
properties have also been studied extensively,
including the relation between Markovianity and
CP-divisible quantum channels \cite{Wolf2008}.

Characteristic functions play an important role in
quantum statistical theory and have been studied for
quantum observables and states \cite{EmoriQCF}.

The goal of the present work is to introduce
characteristic functions associated with quantum
channels themselves. Using the Choi representation
\cite{Choi1975}, we establish a connection between
characteristic-function positivity and complete
positivity of quantum channels.

Applying this framework to time-dependent dynamical
maps yields a characterization of CP-divisibility in
terms of two-time characteristic functions. The
relation between this characterization and information
backflow is also discussed \cite{NakagawaBackflow}. Recent developments have further
explored refined characterizations of non-Markovian
dynamics and their structural properties
\cite{RecentNonMarkov2026,arXiv2601_18822,arXiv2601_12435}.

Finally, we outline the structure of this paper.
In Sec.~II we review quantum dynamical maps and the
Choi representation. In Sec.~III we introduce the
map-dependent quantum characteristic function and
its associated Gram matrix. In Sec.~IV we prove the
Bochner--Choi positivity theorem. In Sec.~V we apply
this framework to time-dependent dynamics and derive
a characterization of CP-divisibility. Numerical
examples for amplitude damping and pure dephasing
models are presented in Sec.~VI. We conclude with a
discussion and outlook in Sec.~VII.

\section{Preliminaries}

\subsection{Quantum dynamical maps}

The evolution of an open quantum system can be described by a dynamical map

\begin{equation}
\Phi(t,0):L(\mathcal H) \rightarrow L(\mathcal H),
\end{equation}

which maps the initial density operator $\rho(0)$ to the state $\rho(t)$.

The intermediate map is defined as

\begin{equation}
\Phi(t,s) = \Phi(t,0)\Phi(s,0)^{-1}.
\end{equation}

A dynamical map is called CP-divisible if the intermediate map $\Phi(t,s)$ is completely positive for all $t \ge s$.

\subsection{Choi representation}

For a linear map $\Phi$ acting on $L(\mathcal H)$, the Choi operator is defined as

\begin{equation}
J(\Phi) = (\Phi \otimes I)(|\Omega\rangle\langle\Omega|),
\end{equation}

where

\begin{equation}
|\Omega\rangle =
\frac{1}{\sqrt{d}}
\sum_i |i\rangle \otimes |i\rangle.
\end{equation}

Choi's theorem states that

\begin{equation}
\Phi \text{ is completely positive}
\quad \Longleftrightarrow \quad
J(\Phi) \ge 0 .
\end{equation}

\section{Map-Dependent Quantum Characteristic Functions}

We now define a characteristic function associated with a quantum channel.

Let

\begin{equation}
\Omega_\Phi = \frac{1}{d} J(\Phi)
\end{equation}

be the normalized Choi operator. The normalization by $1/d$ ensures that $\Omega_\Phi$ has unit trace
for trace-preserving maps, so that it can be interpreted as a
state-like representative of the channel in the Choi space.

Let $\{U_\mu\}$ denote a unitary operator basis on $\mathcal H \otimes \mathcal H$.

The map-dependent quantum characteristic function is defined as

\begin{equation}
\chi_\Phi(U_\mu) =
\mathrm{Tr}(\Omega_\Phi U_\mu).
\end{equation}

The associated Gram matrix is

\begin{equation}
G^\Phi_{\mu\nu}
=
\mathrm{Tr}(
\Omega_\Phi
U_\mu^\dagger U_\nu
).
\end{equation}

This construction resembles the classical characteristic-function framework, but the statistical object is now the dynamical map itself rather than a state.

\section{Bochner--Choi Positivity Theorem}

We now establish the central mathematical result of this work.

\textbf{Theorem 1 (Bochner--Choi positivity theorem).}

Let $\Phi$ be a trace-preserving linear map on $L(\mathcal H)$ and let $\{U_\mu\}$ be a unitary operator basis spanning $L(\mathcal H\otimes\mathcal H)$.
Then

\begin{equation}
\Phi \text{ is completely positive }
\iff
G^\Phi \succeq 0.
\end{equation}

\textit{Proof.}

If $\Phi$ is completely positive, then $J(\Phi)\ge0$ and therefore $\Omega_\Phi\ge0$.

For arbitrary coefficients $c_\mu$, define

\begin{equation}
X = \sum_\mu c_\mu U_\mu .
\end{equation}

Then

\begin{equation}
\sum_{\mu\nu} \bar c_\mu G^\Phi_{\mu\nu} c_\nu
=
\mathrm{Tr}(\Omega_\Phi X^\dagger X).
\end{equation}

Since $\Omega_\Phi \ge 0$ and $X^\dagger X \ge 0$, the trace is non-negative.
Thus $G^\Phi$ is positive semidefinite.

Conversely, suppose $G^\Phi \succeq 0$.
Then

\begin{equation}
\mathrm{Tr}(\Omega_\Phi X^\dagger X)\ge0
\end{equation}

for all operators $X$.
This implies $\Omega_\Phi \ge 0$.
By Choi's theorem, $\Phi$ is completely positive.
\hfill $\square$

\section{Characterization of CP-Divisibility}

For a time-dependent dynamical map we define the two-time characteristic function

\begin{equation}
\chi_{t,s}(U) =
\mathrm{Tr}(\Omega_{t,s}U),
\end{equation}

where

\begin{equation}
\Omega_{t,s} = \frac{1}{d}J(\Phi(t,s)).
\end{equation}

The Gram matrix becomes

\begin{equation}
G^{(t,s)}_{\mu\nu}
=
\mathrm{Tr}(
\Omega_{t,s}
U_\mu^\dagger U_\nu
).
\end{equation}

Since the unitary family $\{U_\mu\}$ spans the full operator space
$L(\mathcal H\otimes\mathcal H)$, the inequality
$\mathrm{Tr}(\Omega_\Phi X^\dagger X)\ge 0$
holds for every operator $X$.
Because every positive operator $P\ge 0$ can be written as
$P=X^\dagger X$, we obtain
$\mathrm{Tr}(\Omega_\Phi P)\ge 0$ for all positive $P$.
Therefore $\Omega_\Phi\ge 0$, and by Choi's theorem
$\Phi$ is completely positive.

\textbf{Theorem 2.}

The dynamical map $\Phi(t,0)$ is CP-divisible if and only if

\begin{equation}
G^{(t,s)} \succeq 0
\quad
\forall t \ge s .
\end{equation}

This follows directly from Theorem 1 applied to the intermediate map $\Phi(t,s)$.

Information backflow has also been analyzed in terms
of time-local generators and their relation to
divisibility properties of dynamical maps
\cite{NakagawaBackflow}.

\section{Numerical Examples}

\subsection{Amplitude damping}

We consider the time-local master equation

\begin{equation}
\dot{\rho}(t)
=
\gamma(t)
(\sigma_- \rho \sigma_+ -\tfrac12\{\sigma_+\sigma_-,\rho\}).
\end{equation}

The decay rate

\begin{equation}
\gamma(t)=\gamma_0 + a e^{-t}\cos(\omega t)
\end{equation}

induces temporary non-Markovian behavior.

The survival probability is

\begin{equation}
\eta(t)=\exp\left(-\int_0^t \gamma(\tau)d\tau\right).
\end{equation}

The intermediate map parameter is

\begin{equation}
r(t,s)=\frac{\eta(t)}{\eta(s)}.
\end{equation}

Values exceeding unity signal non-CP-divisible dynamics.

\begin{figure}
\includegraphics[width=\linewidth]{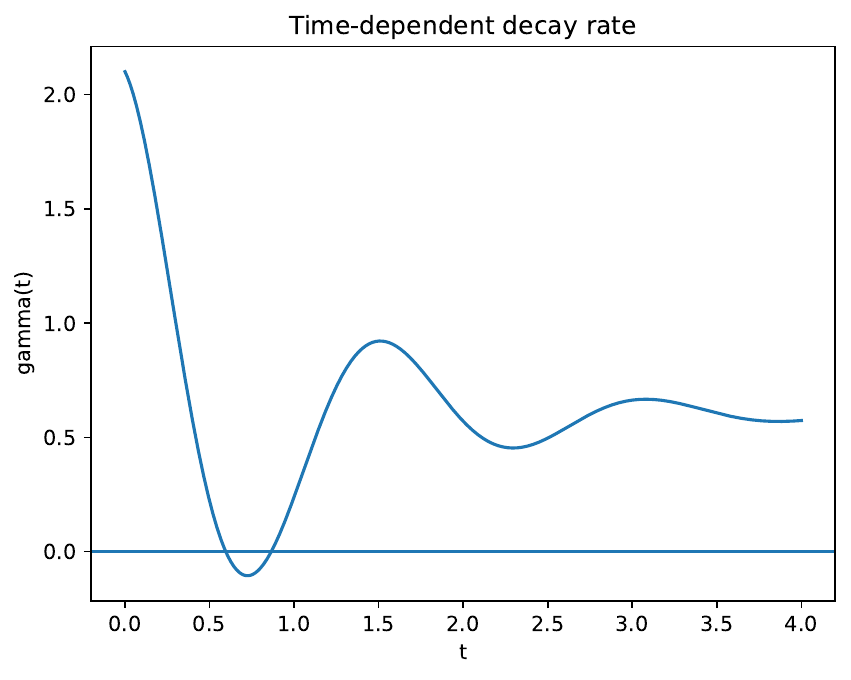}
\caption{Time-dependent decay rate.}
\label{fig:gamma}
\end{figure}

Figure~\ref{fig:gamma} shows the time-dependent decay rate
$\gamma(t)$ used in the amplitude-damping model.
Negative intervals of $\gamma(t)$ correspond to
non-Markovian dynamics.

\subsection{Pure dephasing}

We also consider a pure dephasing model

\begin{equation}
\dot\rho(t)=\gamma(t)
(\sigma_z\rho\sigma_z-\rho).
\end{equation}

Negative intervals of $\gamma(t)$ lead to coherence revival and non-Markovian dynamics.

\begin{figure}
\includegraphics[width=\linewidth]{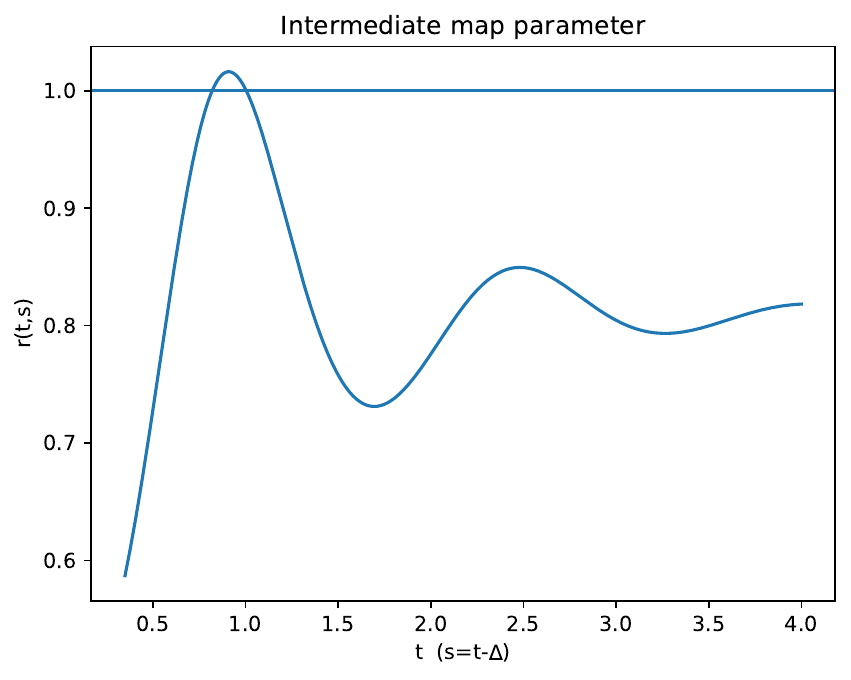}
\caption{Intermediate map parameter $r(t,s)$.}
\label{fig:rts}
\end{figure}

The intermediate map parameter $r(t,s)=\eta(t)/\eta(s)$
is plotted in Fig.~\ref{fig:rts}.  
Values exceeding unity indicate that the intermediate map
is not completely positive and therefore signal a breakdown
of CP-divisibility.

\begin{figure}
\includegraphics[width=\linewidth]{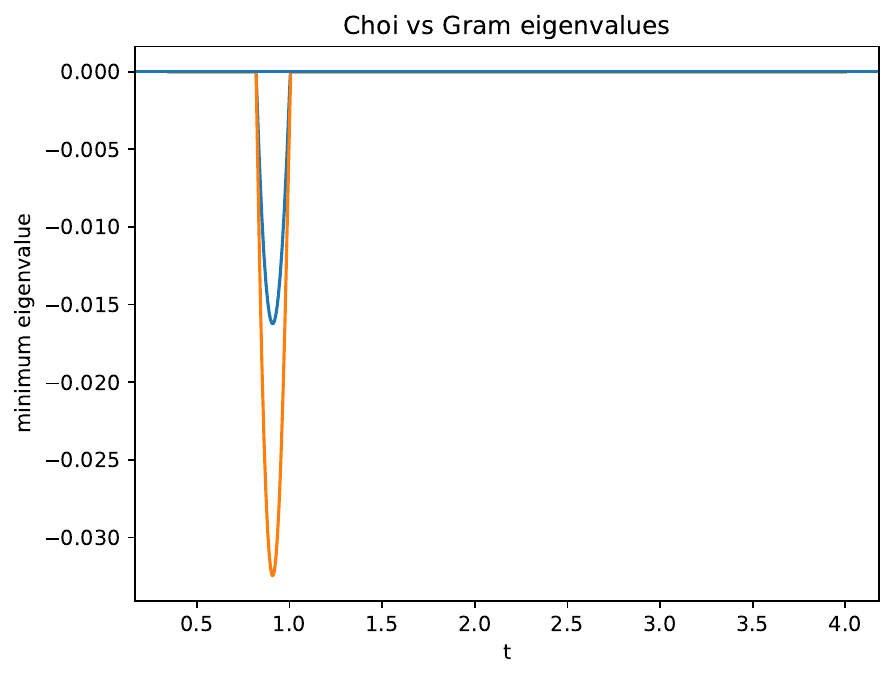}
\caption{Minimum eigenvalues of the Choi operator and the Gram matrix.}
\label{fig:gram}
\end{figure}

The smallest eigenvalues of the Choi operator and
the Gram matrix are shown in Fig.~\ref{fig:gram}.
The negativity of the Gram matrix coincides with
the negativity of the Choi eigenvalues, confirming
the Bochner--Choi theorem.

\begin{figure}
\includegraphics[width=\linewidth]{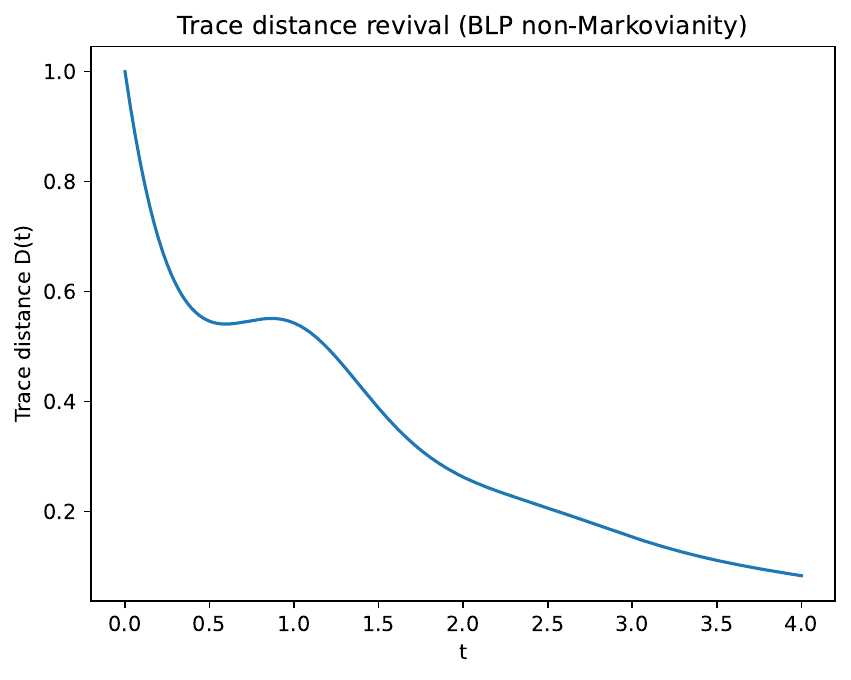}
\caption{
Trace distance between two states.
Revival of the trace distance indicates
information backflow.
}
\label{fig:BLP}
\end{figure}

\begin{figure}
\includegraphics[width=\linewidth]{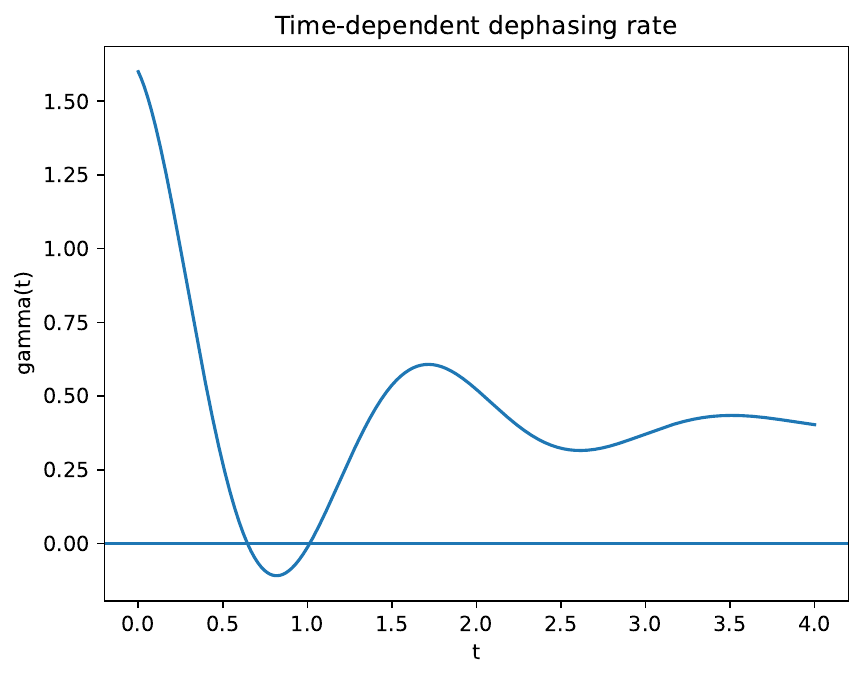}
\caption{Time dependent dephasing rate.}
\end{figure}

\begin{figure}
\includegraphics[width=\linewidth]{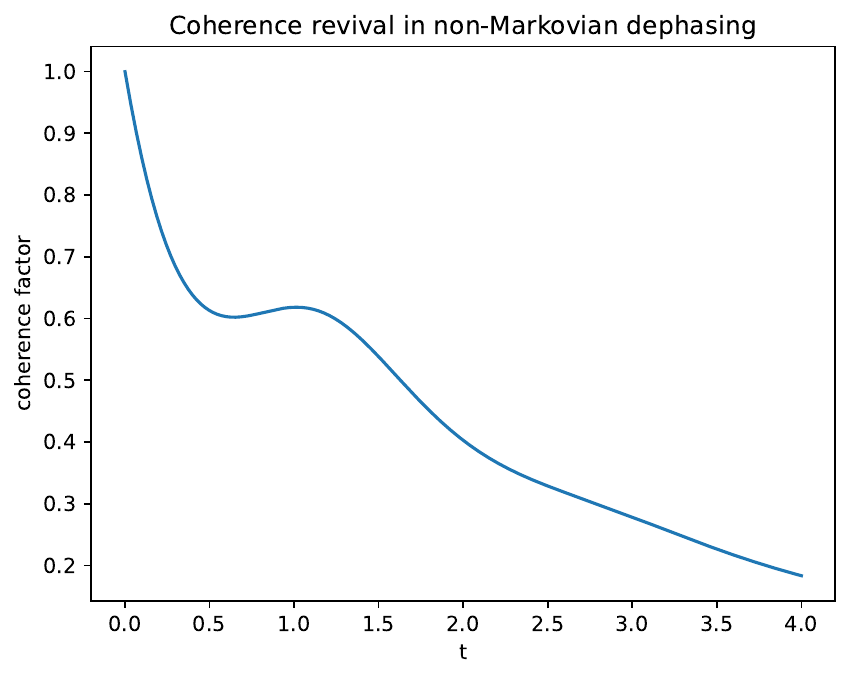}
\caption{Coherence revival in non-Markovian dephasing.}
\end{figure}

\subsection{Trace distance revival}

To illustrate information backflow we compute the
trace distance between two states

\[
D(t)=\frac12\|\rho_1(t)-\rho_2(t)\|_1 .
\]

For amplitude damping one obtains

\[
D(t)=\eta(t).
\]

Revival of $D(t)$ therefore occurs whenever
$\gamma(t)<0$, corresponding to the BLP
non-Markovianity measure \cite{Breuer2009}.

\section{Discussion}

The present framework introduces characteristic functions defined for dynamical maps rather than quantum states.

The Bochner--Choi theorem shows that positivity of the associated Gram matrix characterizes complete positivity of quantum channels.

Although Theorem 1 is basis independent, numerical implementation
requires a concrete operator basis. In the present work we use a
unitary basis, such as the Pauli basis for qubit models, because it
provides a compact and numerically stable representation of the Gram matrix.

From a practical viewpoint, reconstruction of the map-dependent
characteristic function from experimental data will be affected by
statistical noise. In such situations, positivity of the Gram matrix
should be interpreted in a robust sense, for example through
confidence intervals on its smallest eigenvalue.
For larger Hilbert-space dimension $d$, the size of the Gram matrix
grows with the operator basis, so efficient basis truncation or
tensor-network-inspired compression may become important.

This provides a new perspective on non-Markovian quantum dynamics.
In particular, negativity of the Gram matrix signals the breakdown of CP-divisibility and is closely related to information backflow.

Future work may extend the framework to multi-time processes and process tensors.

Recent works have also investigated refined structural
and operational aspects of non-Markovian dynamics,
which may provide further context for the present
characteristic-function-based approach \cite{RecentNonMarkov2026,arXiv2601_18822,arXiv2601_12435}.


\bibliographystyle{apsrev4-2}
\bibliography{references}

\end{document}